\documentclass[a4paper,12pt]{article}
\usepackage[T1]{fontenc}
\usepackage{amsfonts}
\usepackage{ae,aecompl}
\usepackage{amsmath}
\usepackage{amssymb}
\usepackage{amsthm}
\usepackage[pdftex]{graphicx}
\usepackage{pstricks}
\usepackage{tikz}
 \usepackage{epstopdf}
\usepackage[small]{caption}   
\usepackage{color}
\usepackage{enumerate} 
\usepackage{floatrow}
\usepackage{hyperref}
\usepackage{mathdots}
\usepackage{hyperref}
\hypersetup{
    colorlinks=true,
    linkcolor=blue,
    filecolor=magenta,      
    urlcolor=blue,
}

\newcommand{\ie}[0]{\textit{i.e.}}

\definecolor{lightgray}{gray}{0.8}

\usepackage{media9}

\title{Tomographic X-ray data of a cross phantom}
\author{S. Latva-\"{A}ij\"{o}\footnote{Department of Mathematics and Statistics, University of Helsinki, Finland (salla.latva-aijo@helsinki.fi)}, 
A. Meaney\footnote{Department of Mathematics and Statistics, University of Helsinki, Finland (alexander.meaney@helsinki.fi)},
\ and S. Siltanen\footnote{Department of Mathematics and Statistics, University of Helsinki, Finland (samuli.siltanen@helsinki.fi)}}

\begin{document}

\maketitle

\abstract{This is the documentation of the tomographic X-ray data of a dynamic cross phantom made available \href{%https://zenodo.org/record/1183532#.WpA35Y5rIy1}{here}. 
The data can be freely used for scientific purposes with appropriate references to the data and to this document in \url{http://arxiv.org/}. The data set consists of (1) the X-ray sinogram with 16 or 30 time frames (depending on the resolution) of 2D slices of the phantom and (2) the corresponding static and dynamic measurement matrices modeling the linear operation of the X-ray transform. Each of these sinograms was obtained from a measured 360-projection fan-beam sinogram by down-sampling and taking logarithms.

\section{Introduction}

The main idea behind the project was to create real CT measurement data for testing sparse data and dynamic data tomography algorithms. 
A simple target which is changing in time was prepared by crossing aluminum stick and graphite stick so that sticks are in slightly different angles compared to each other.  The system was stabilized using candle wax. See Figure~\ref{fig:SetupAndProjections} of the preparation phase of the phantom. When the target rotates and we observe it along the z-axis, the cross sections of the sticks move toward each other or appart from each other in every time step. The time-dependent target is challenge for sparse dynamic tomography applications. The CT data in this data set has been used for example with the modified level set reconstruction method, see \href{%https://blog.fips.fi/}{FIPS Computational Blog}.

\section{Contents of the data set}\label{sec:datasets}

The data set contains the following MATLAB\footnote{MATLAB is a registered trademark of The MathWorks, Inc.} data files:
\begin{itemize}
\item  {\tt DataStatic\_128x60.mat},
\item  {\tt DataStatic\_128x15.mat},
\item  {\tt DataDynamic\_128x60.mat},
\item  {\tt DataDynamic\_128x15.mat},
\item  {\tt DataStatic\_256x60.mat},
\item  {\tt DataStatic\_256x15.mat},
\item  {\tt DataDynamic\_256x60.mat},
\item  {\tt DataDynamic\_256x15.mat},
\item {\tt FullSizeSinogram.mat} and
\item  {\tt GroundTruthReconstruction.mat}.
\end{itemize}
The first four of these files contain CT sinogram and the corresponding measurement matrix with the resolution $128 \times 128$ as spatial resolution and $16$ as a temporal resolution in 3D. In the last four files the contents are the same, but spatial resolution is $256 \times 256$ and temporal resolution in 3D is $30$ time frames. The word "{\tt Static}" in the file name means that the data leads to reconstructions with the same geometry in every time instance.  The word "{\tt Dynamic}" in the file name means that the projection angles shift by one degree in every time step, for example, if the angles for time $t_1$ are $[1\quad13\quad25 \,\cdots]$, then for time $t_2$, the angles are $[2\quad14\quad26 \,\cdots]$. The detailed contents of every file are listed below.

\bigskip\noindent
{\tt DataStatic\_128x60.mat} contains the following variables:
\begin{enumerate}
\item Sparse matrix {\tt A} of size $134\,400\times 262\,144$ sparse double; measurement matrix.
\item Matrix {\tt sinogram} of size $140\times 960$; sinogram (60 projections out of full 360 degree circle).
\end{enumerate}

\bigskip\noindent
{\tt DataStatic\_128x15.mat} contains the following variables:
\begin{enumerate}
\item Sparse matrix {\tt A} of size $33\,600\times 262\,144$; measurement matrix.
\item Matrix {\tt sinogram} of size $140\times 240$; sinogram (15 projections).
\end{enumerate}

\bigskip\noindent
{\tt DataDynamic\_128x60.mat} contains the following variables:
\begin{enumerate}
\item Sparse matrix {\tt A} of size $134\,400\times 262\,144$; measurement matrix.
\item Matrix {\tt m} of size $140\times 960$; sinogram (60 projections).
\end{enumerate}

\bigskip\noindent
{\tt DataDynamic\_128x15.mat} contains the following variables:
\begin{enumerate}
\item Sparse matrix {\tt A} of size $33\,600\times 252\,144$; measurement matrix.
\item Matrix {\tt sinogram} of size $140\times 240$; sinogram (15 projections).
\end{enumerate}

\bigskip\noindent
{\tt DataStatic\_256x60.mat} contains the following variables:
\begin{enumerate}
\item Sparse matrix {\tt A} of size $504\,000\times 1\,966\,080$; measurement matrix.
\item Matrix {\tt sinogram} of size $280\times 1800$; sinogram (60 projections).
\end{enumerate}

\bigskip\noindent
{\tt DataStatic\_256x15.mat} contains the following variables:
\begin{enumerate}
\item Sparse matrix {\tt A} of size $126\,000\times 1\,966\,080$; measurement matrix.
\item Matrix {\tt sinogram} of size $280\times 450$; sinogram (15 projections).
\end{enumerate}

\bigskip\noindent
{\tt DataDynamic\_256x60.mat} contains the following variables:
\begin{enumerate}
\item Sparse matrix {\tt A} of size $504\,000\times 1\,966\,080$; measurement matrix.
\item Matrix {\tt sinogram} of size $280\times 1800$; sinogram (60 projections).
\end{enumerate}

\bigskip\noindent
{\tt DataDynamic\_256x15.mat} contains the following variables:
\begin{enumerate}
\item Sparse matrix {\tt A} of size $126\,000\times 1\,966\,080$; measurement matrix.
\item Matrix {\tt sinogram} of size $280\times 450$; sinogram (15 projections).
\end{enumerate}

\bigskip\noindent
{\tt FullSizeSinogram.mat} contains the following variables:
\begin{enumerate}
\item Matrix {\tt sinogram} of size $2240\times 360$; original (measured) sinogram
of 360 projections.
\end{enumerate}

\bigskip\noindent
{\tt GroundTruthReconstruction.mat} contains the following variables:
\begin{enumerate}
\item Matrix {\tt GroundTruthReconstruction.mat} of size $2240\times 360$; a high-resolution filtered backprojection
reconstruction computed from the larger sinogram of 360 projections of the cross phantom ("ground truth"). See Figure~\ref{fig:GroundTruthReco}. 
\end{enumerate}

\bigskip\noindent
Details on the X-ray measurements are described in Section \ref{sec:Measurements} below.

The model for the CT problem is
\begin{equation}\label{eqn:Axm}
 {\tt A*x=m(:)},
\end{equation}
where {\tt m(:)} denotes the standard vector form of matrix {\tt m} in MATLAB ({\tt m} corresponds to matrix {\tt sinogram}) and {\tt x} is the reconstruction in vector form. In other words, the reconstruction task is to find a vector {\tt x} that (approximately) satisfies \eqref{eqn:Axm} and possibly also meets some additional regularization requirements.
A demonstration of the use of the data is presented in Section \ref{sec:Demo} below.

\section{X-ray measurements}\label{sec:Measurements}
The data in the sinogram is X-ray tomographic (CT) data of a 2D cross-section of the dynamic phantom built from sticks and candle wax and measured with a custom built CT device shown in Figure~\ref{fig:CTmachine}. 
\begin{itemize}
\item The X-ray tube is a model XTF5011 manufactured by Oxford Instruments. This model is no longer sold by Oxford Instruments, although they have newer, similar models available. The tube uses a molybdenum ($Z = 42$) target. 
\item The rotation stage is a Thorlabs model CR1/M-27.
\item The flat panel detector is a Hamamatsu Photonics C7942CA-22. The active area of the 
at panel detector is $120$ mm $\times 120$ mm. It consists of a $2400 \times 2400$ array of $50$ $\mu$m pixels. According to the manufacturer, the number of active pixels is  $2240 \times 2344$. However, the image files actually generated by the camera were $2240 \times 2368$ pixels in size.
\end{itemize}

\begin{figure}[h]
\begin{picture}(390,300)
\put(0,0){\includegraphics[width=400pt]{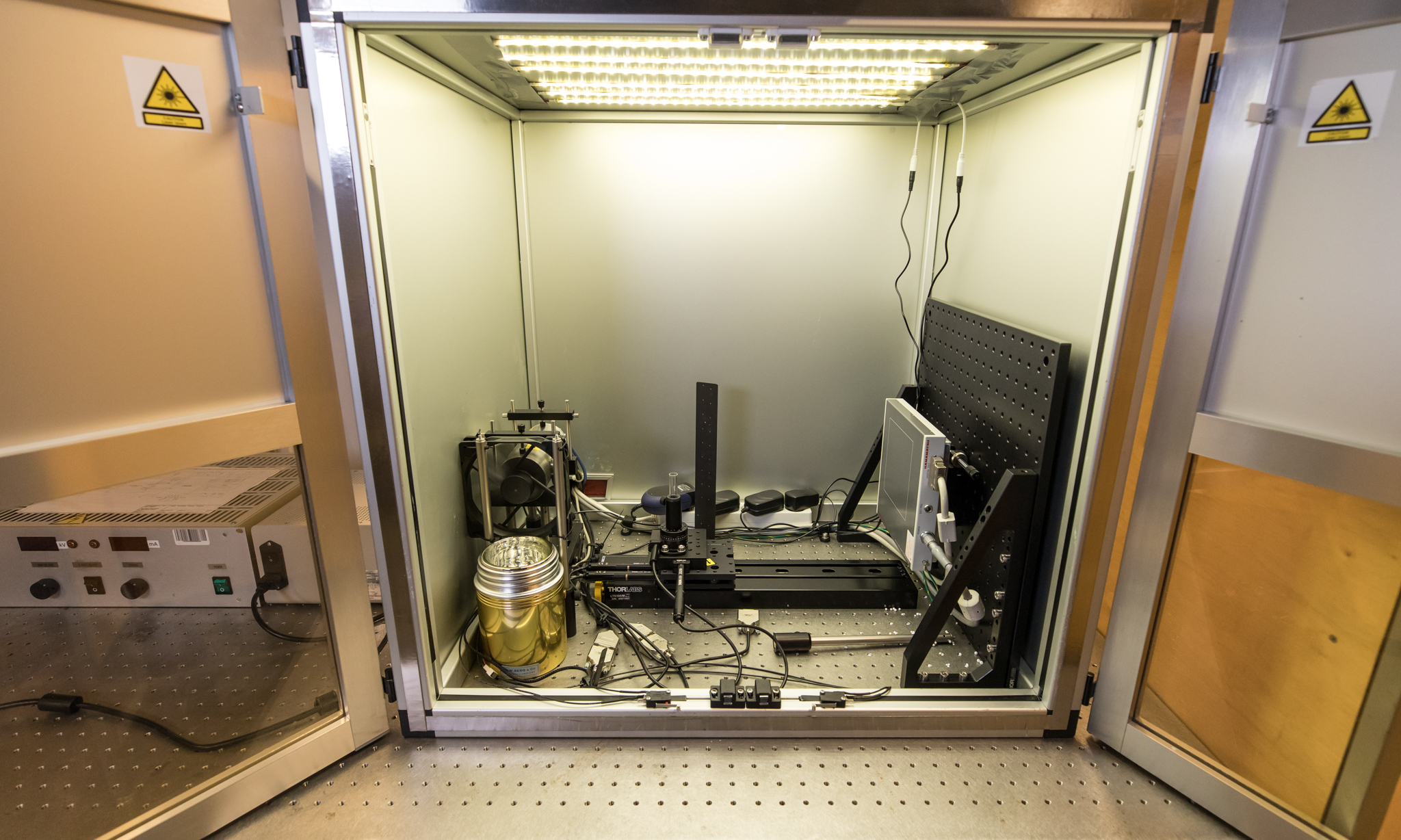}}
\end{picture}
\caption{The custom-made measurement device at University of Helsinki. Figure was taken by Markus Juvonen.}\label{fig:CTmachine}
\end{figure}
The measurement setup was assembled in 2015 by Alexander Meaney as an MSc thesis project \cite{meaney2015design}. The setup is illustrated in Figure~\ref{fig:CTmachine} and the measurement geometry is shown in Figure~\ref{k1}. A set of 360 cone-beam projections with resolution $2240 \times 2368$ and the angular step of one (1) degree was measured. The exposure time was 1000 ms (i.e., one second). The X-ray tube acceleration voltage was 50 kV and tube current 0.9 mA. See Figure~\ref{fig:SetupAndProjections} for example of the resulting projection image. 

From the 2D projection images, the 230 rows from the projection image were selected to form sinograms of the dynamic phantom of resolution $2240 \times 360$. These sinograms were further down-sampled by binning, taken logarithms and normalizing to obtain the {\tt sinogram} in all the files specified in Section~\ref{sec:datasets}. The organization of the pixels in the sinograms and the reconstructions is illustrated in Figure~\ref{fig:pixelDemo}.

\begin{figure}[h]
\includegraphics[width = 4.7cm]{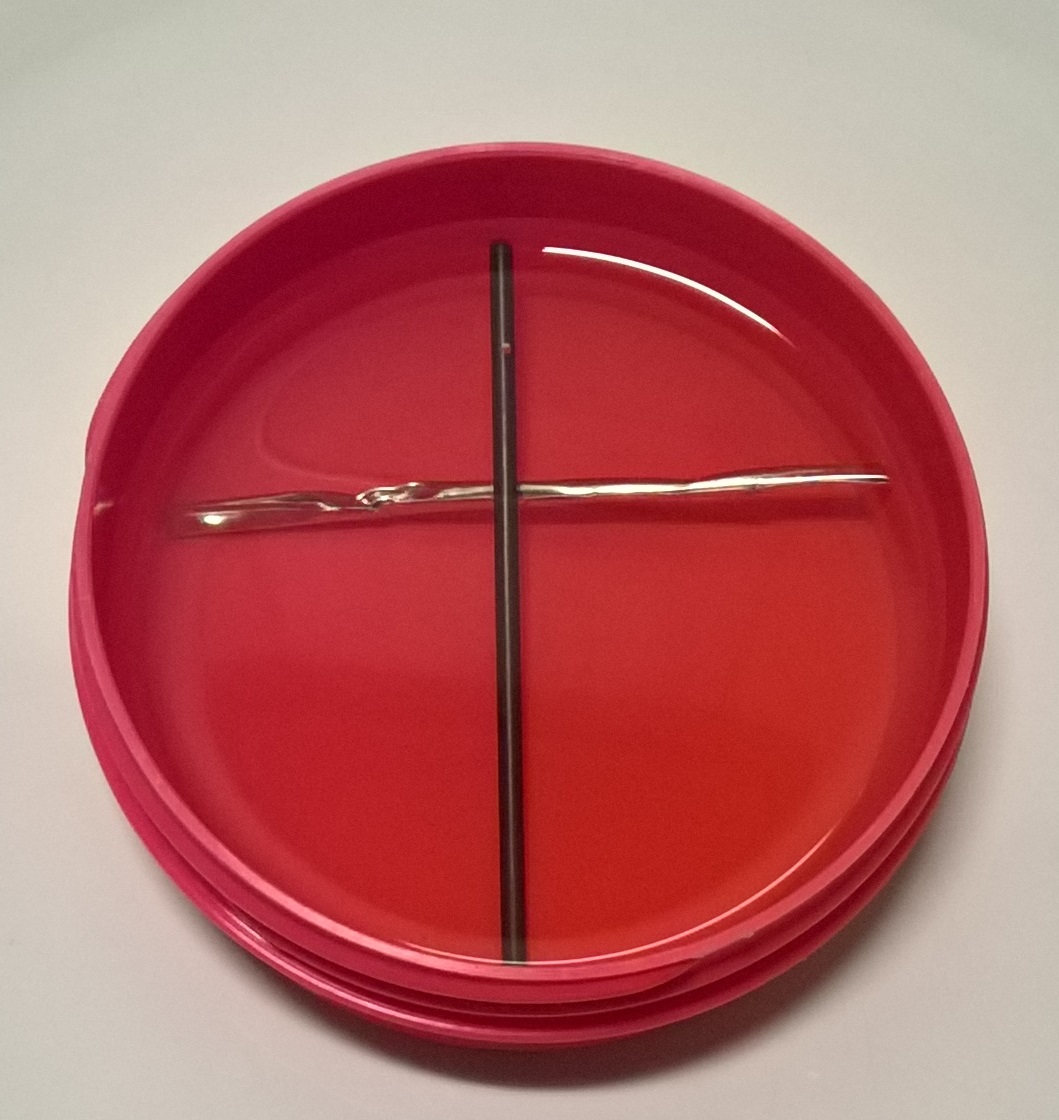}
\includegraphics[width = 5.7cm]{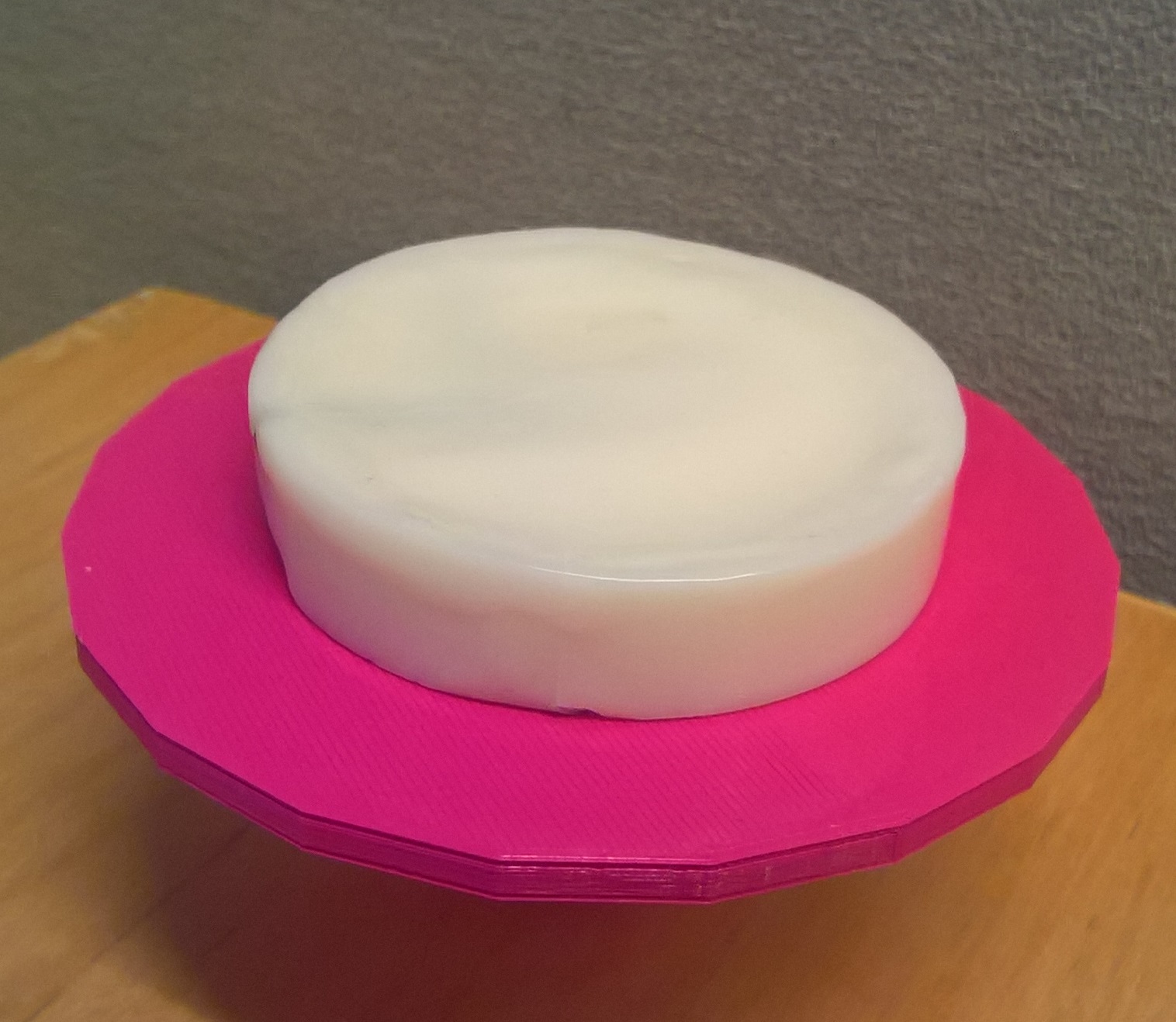}
\includegraphics[height=141pt]{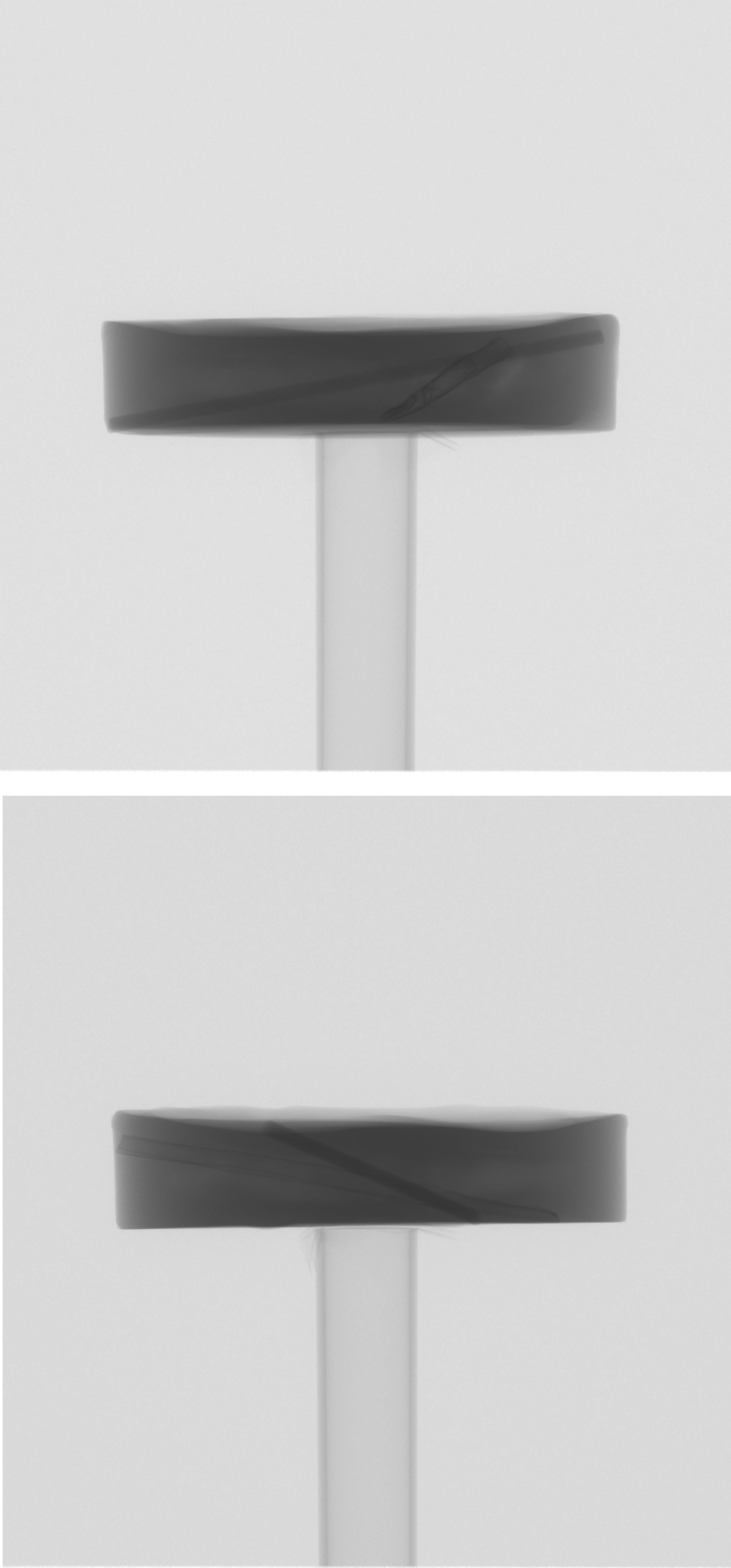}
\caption{\emph{Left}: The phantom in the preparation phase, when the positions of the sticks were set. The darker stick is a graphite taken from a pen. The aluminum stick was made by rolling aluminum sheet, which explains its irregular shape. It is also hollow and contains air, which can been seen in the resulting reconstructions. \emph{Middle}: The ready-made phantom on the sample holder, which can be attached to a computer-controlled rotator platform on the CT device, see Figure~\ref{fig:CTmachine}. \emph{Right}: Two examples of the resulting 2D projection images.} \label{fig:SetupAndProjections} 
\end{figure}

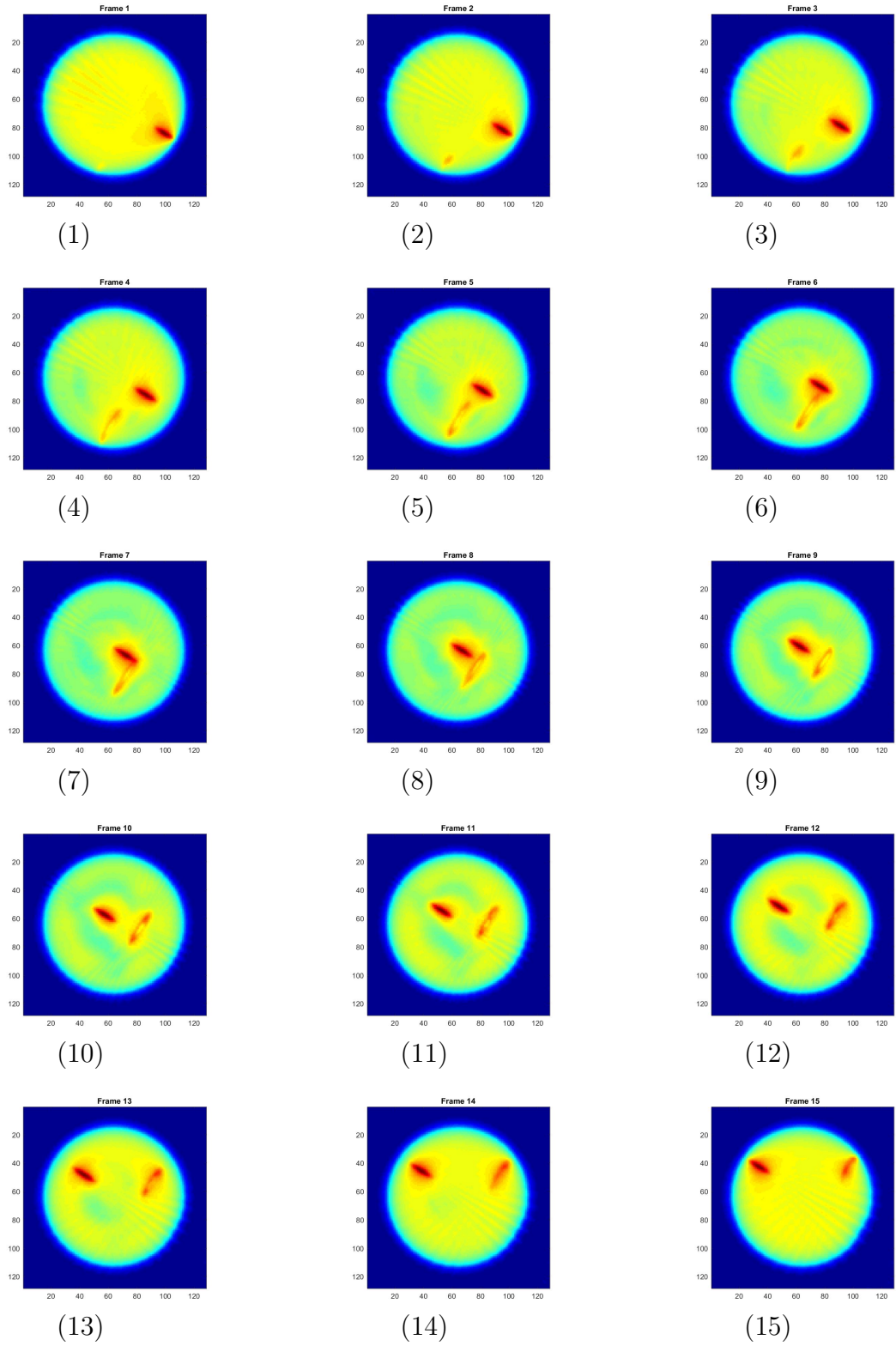
\begin{figure}
\begin{picture}(390,500)
\put(0,500){\includegraphics[width=130pt]{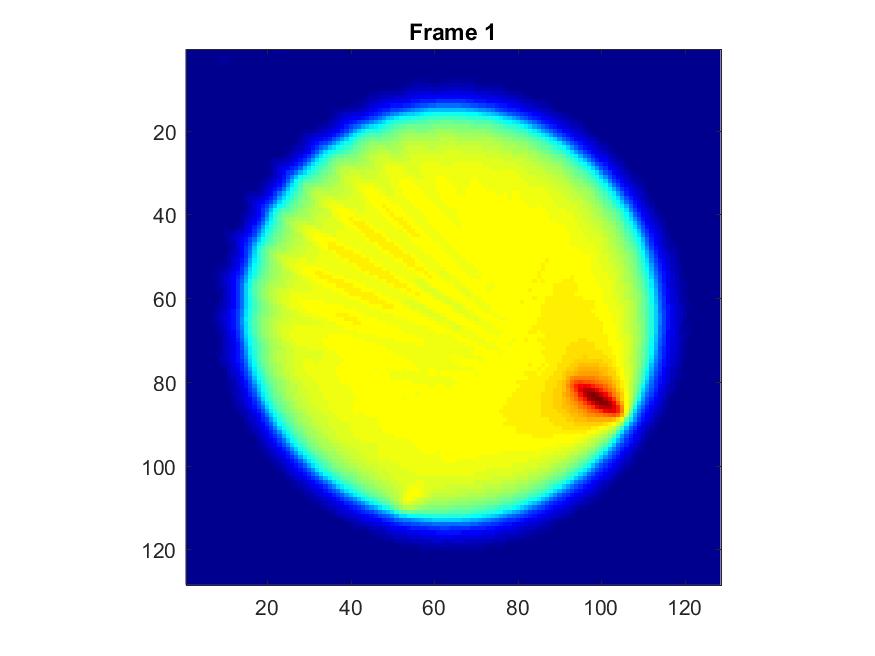}}
\put(150,500){\includegraphics[width=130pt]{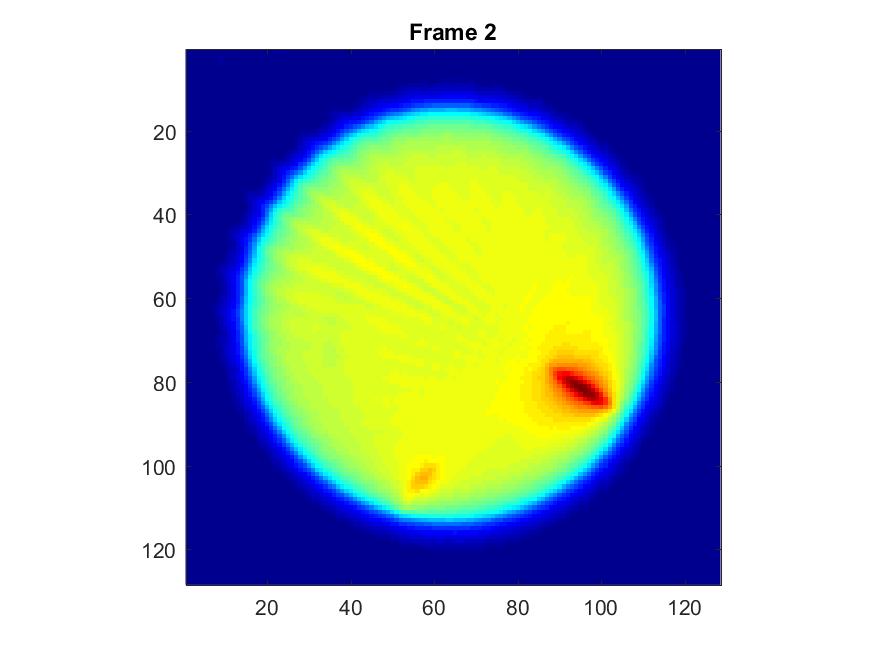}}
\put(300,500){\includegraphics[width=130pt]{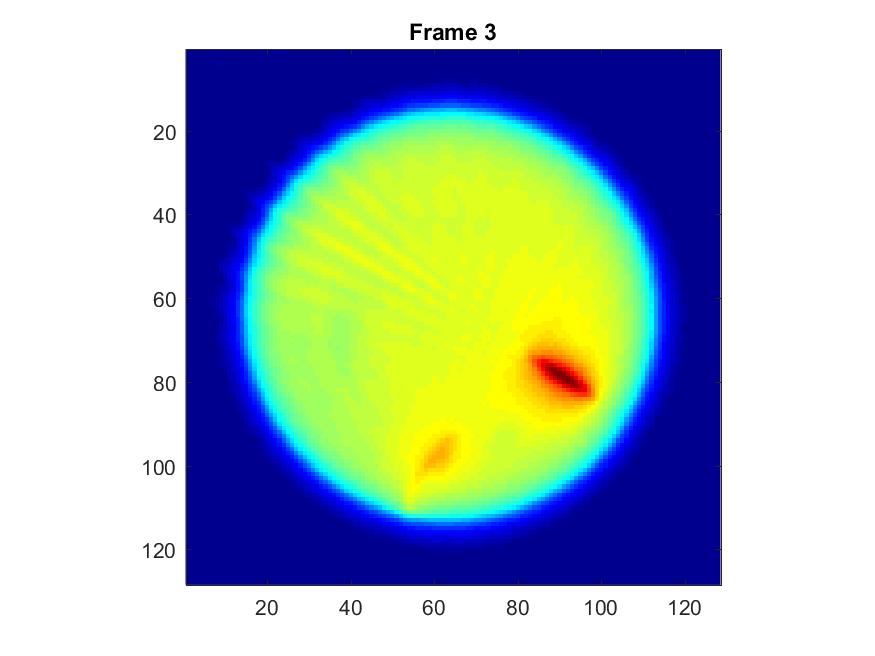}}
\put(0,380){\includegraphics[width=130pt]{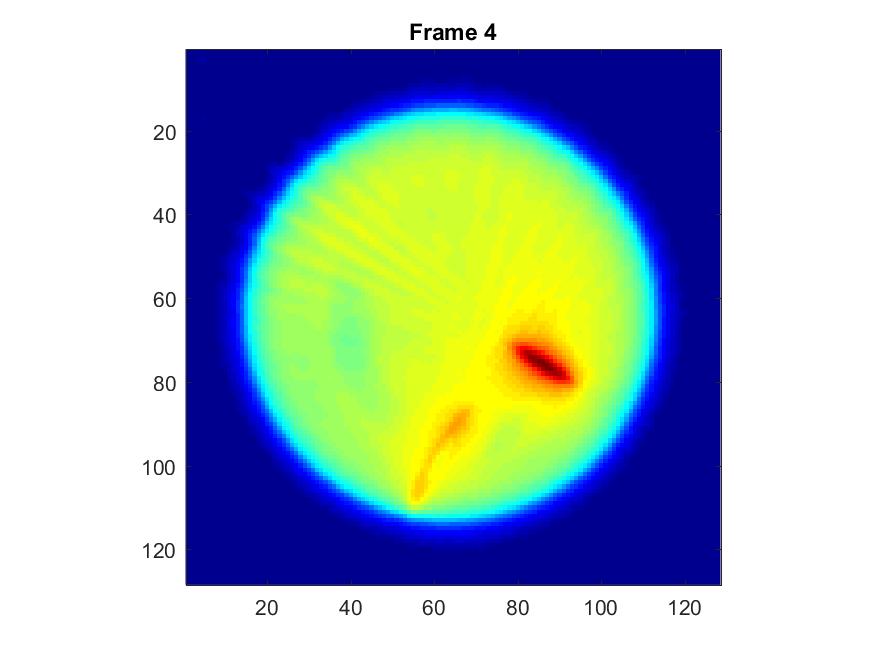}}
\put(150,380){\includegraphics[width=130pt]{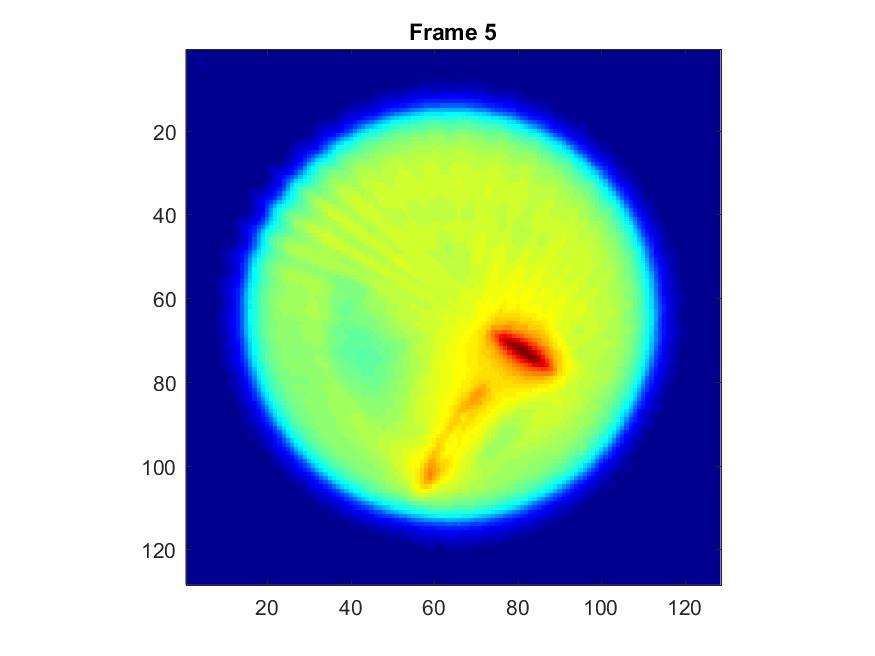}}
\put(300,380){\includegraphics[width=130pt]{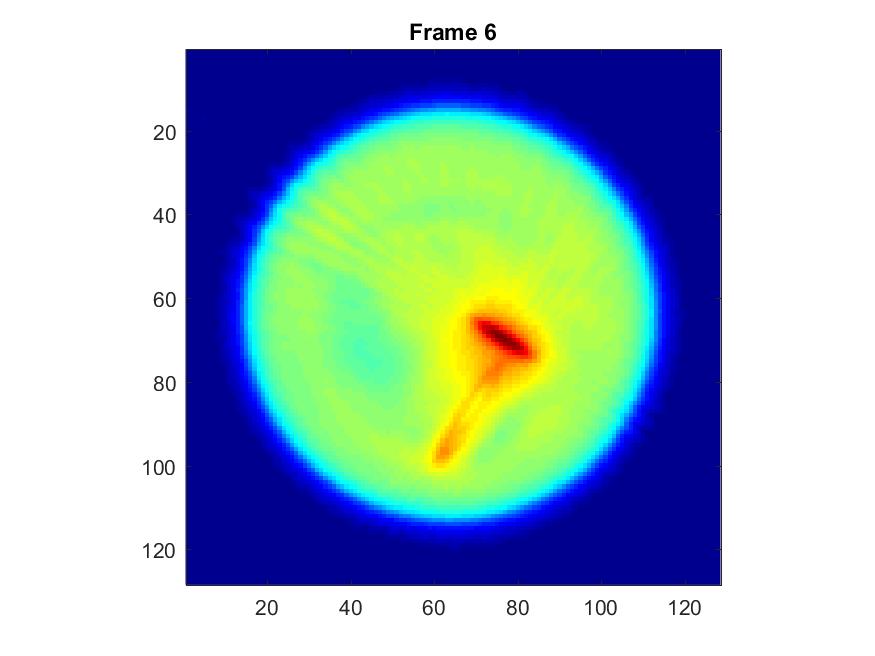}}
\put(0,260){\includegraphics[width=130pt]{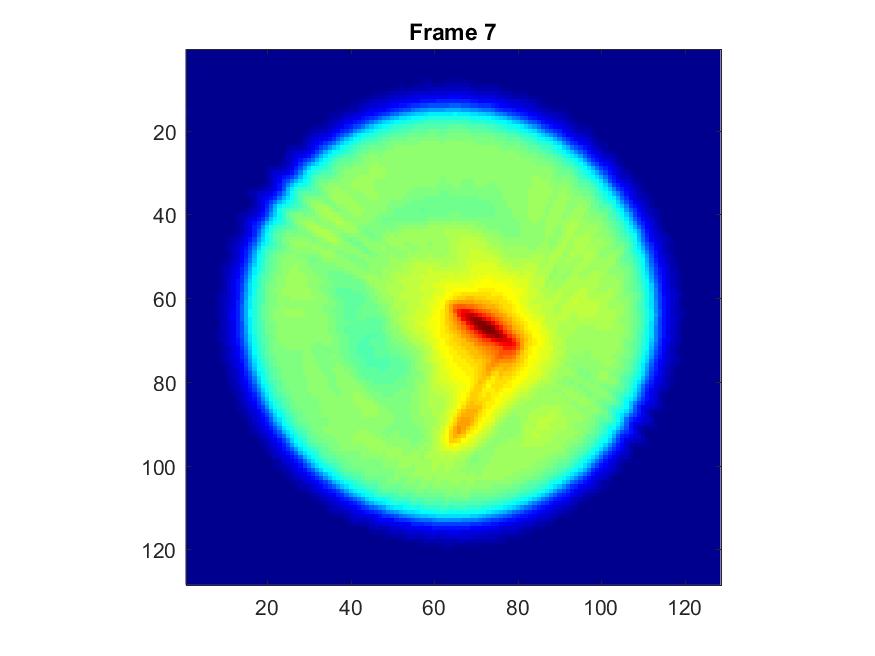}}
\put(150,260){\includegraphics[width=130pt]{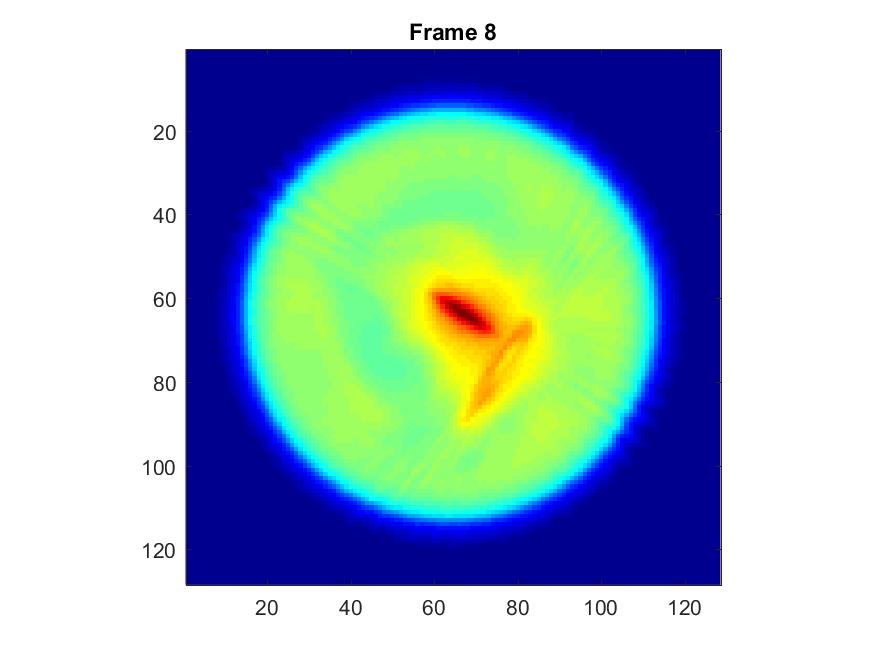}}
\put(300,260){\includegraphics[width=130pt]{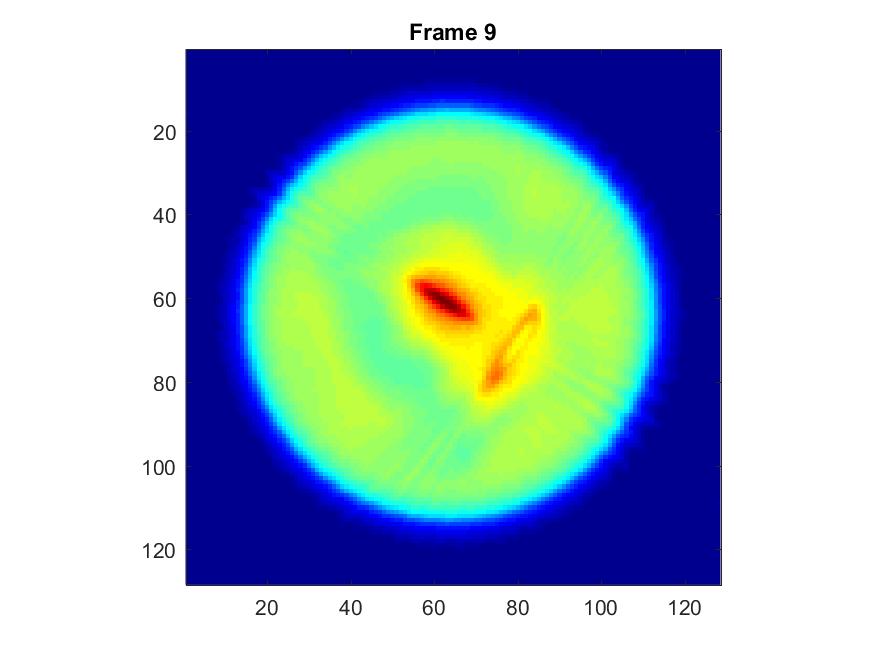}}
\put(0,140){\includegraphics[width=130pt]{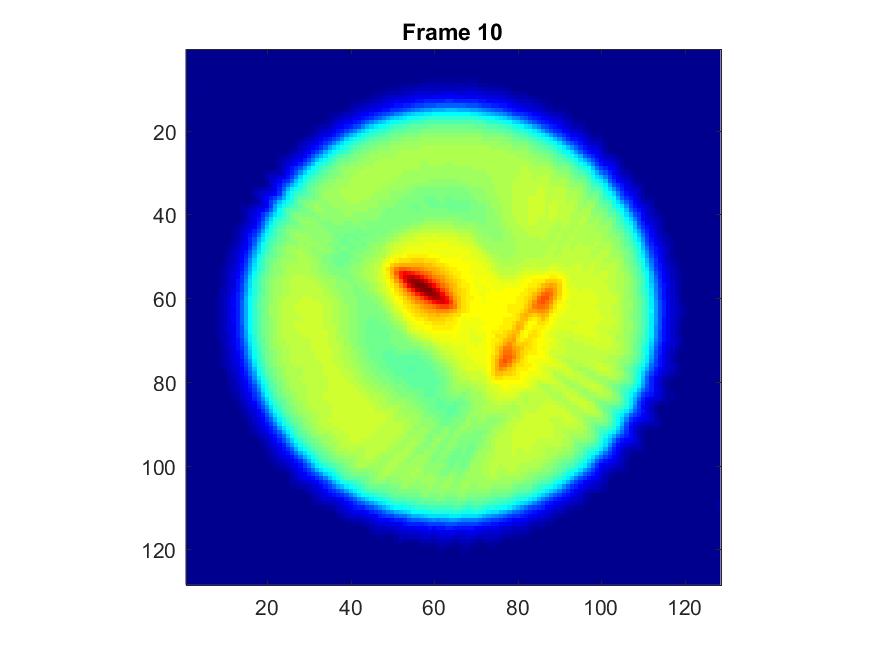}}
\put(150,140){\includegraphics[width=130pt]{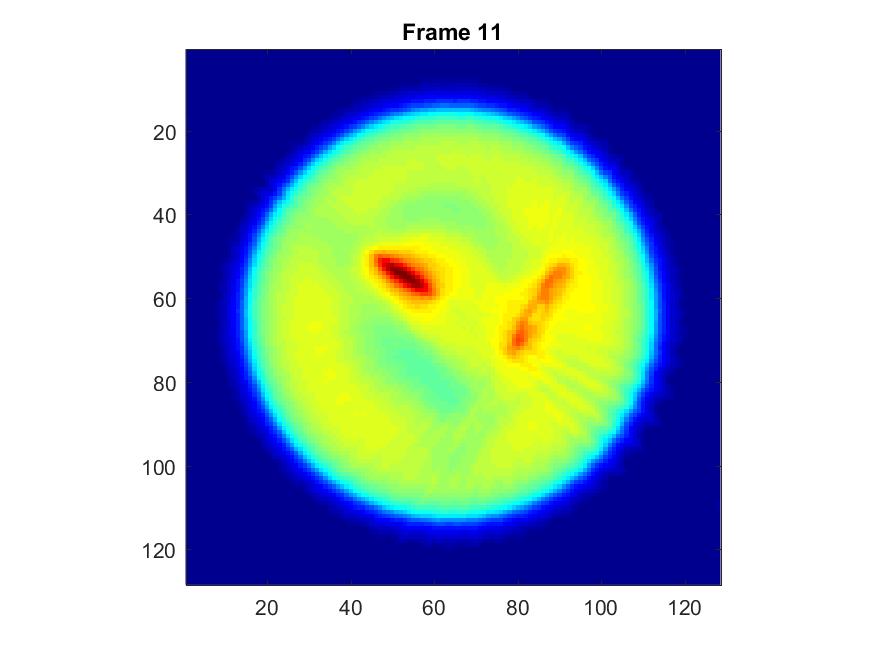}}
\put(300,140){\includegraphics[width=130pt]{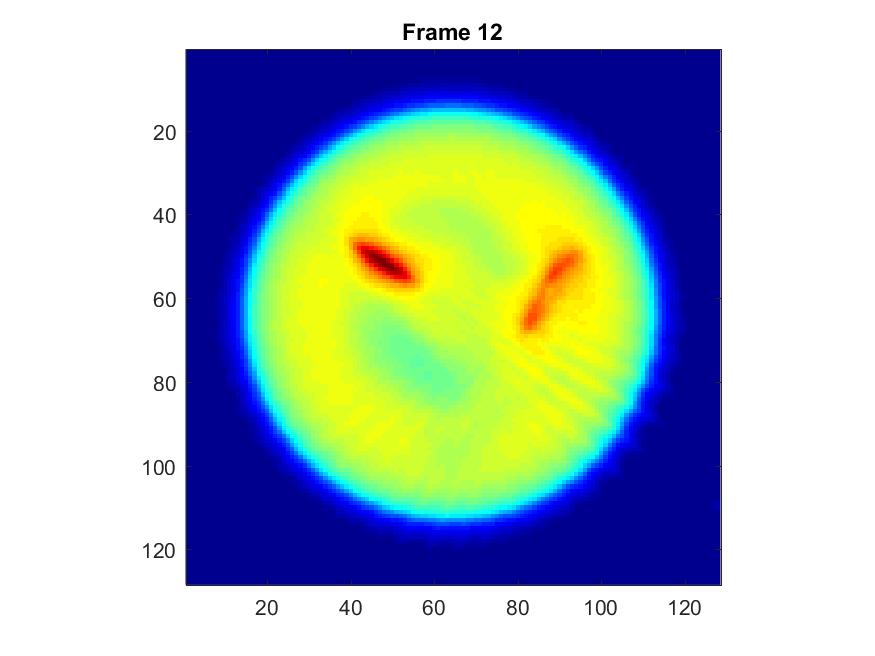}}
\put(0,20){\includegraphics[width=130pt]{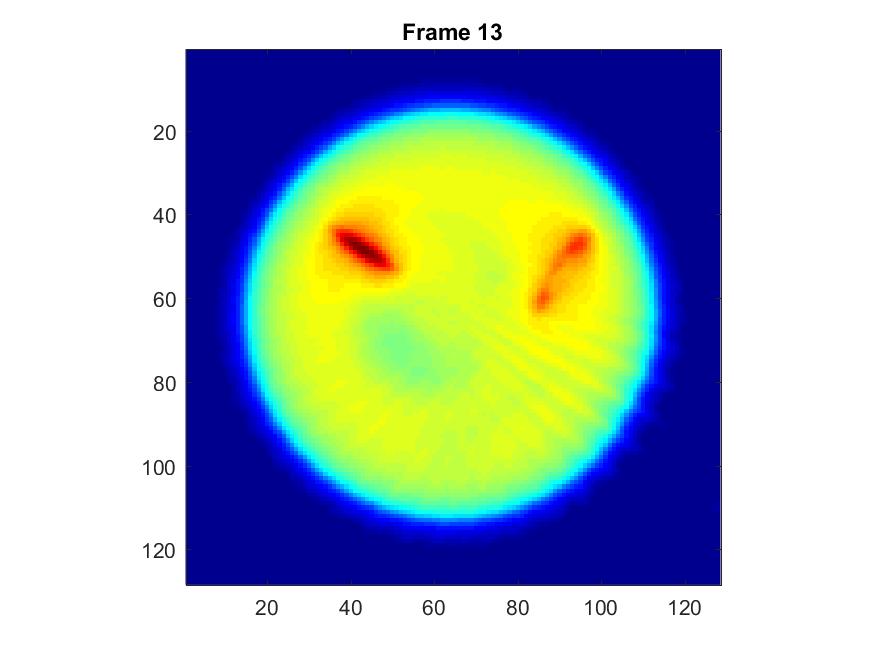}}
\put(150,20){\includegraphics[width=130pt]{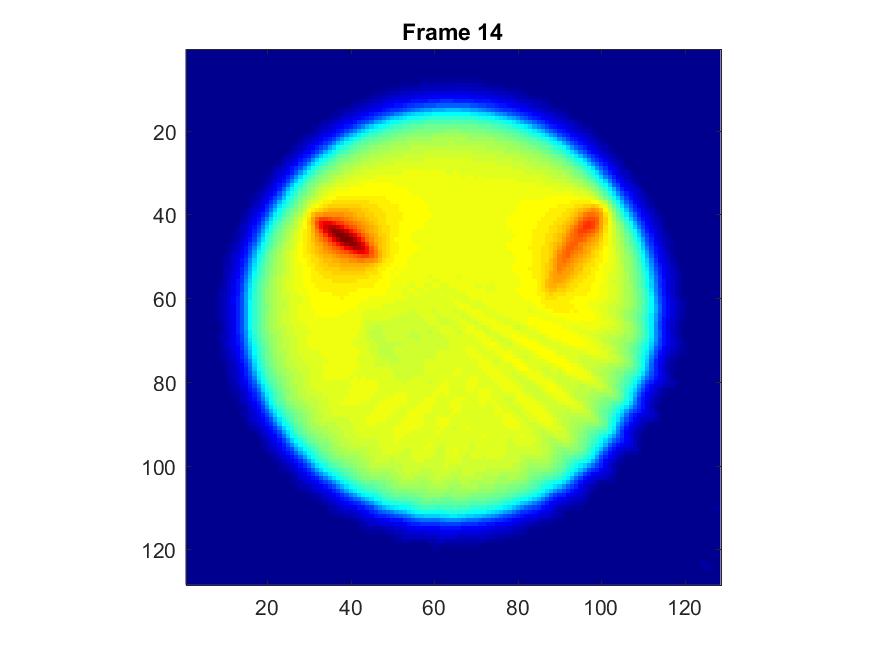}}
\put(300,20){\includegraphics[width=130pt]{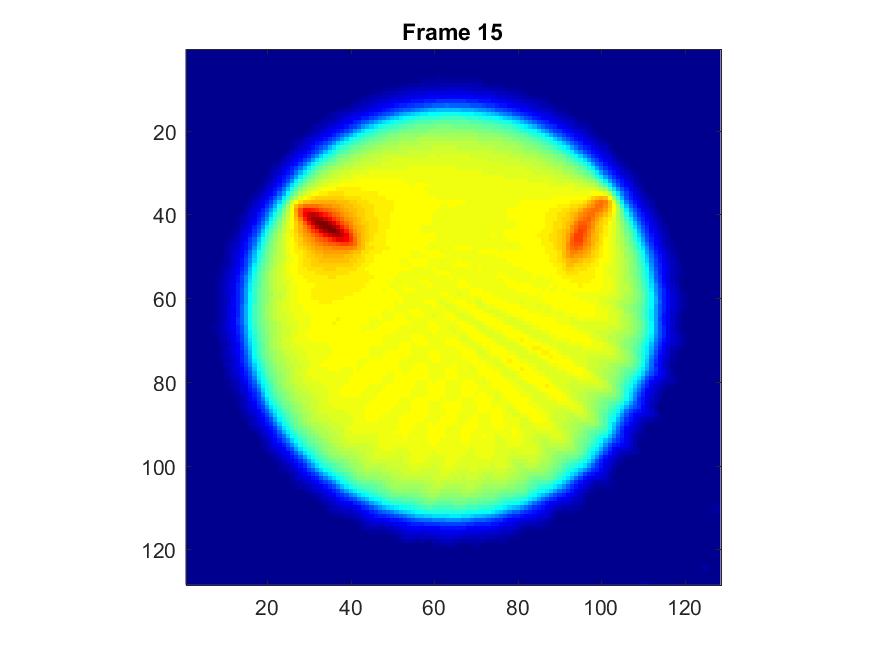}}
\put(42,490){(1)}
\put(192,490){(2)}
\put(342,490){(3)}
\put(42,370){(4)}
\put(192,370){(5)}
\put(342,370){(6)}
\put(42,250){(7)}
\put(192,250){(8)}
\put(342,250){(9)}
\put(42,130){(10)}
\put(192,130){(11)}
\put(342,130){(12)}
\put(42,10){(13)}
\put(192,10){(14)}
\put(342,10){(15)}
\end{picture}
\caption{Tikhonov reconstructions of 15 different example frames showing the movement of the cross sections of the aluminum and graphite sticks. }\label{fig:TargetPhantom}
\end{figure}

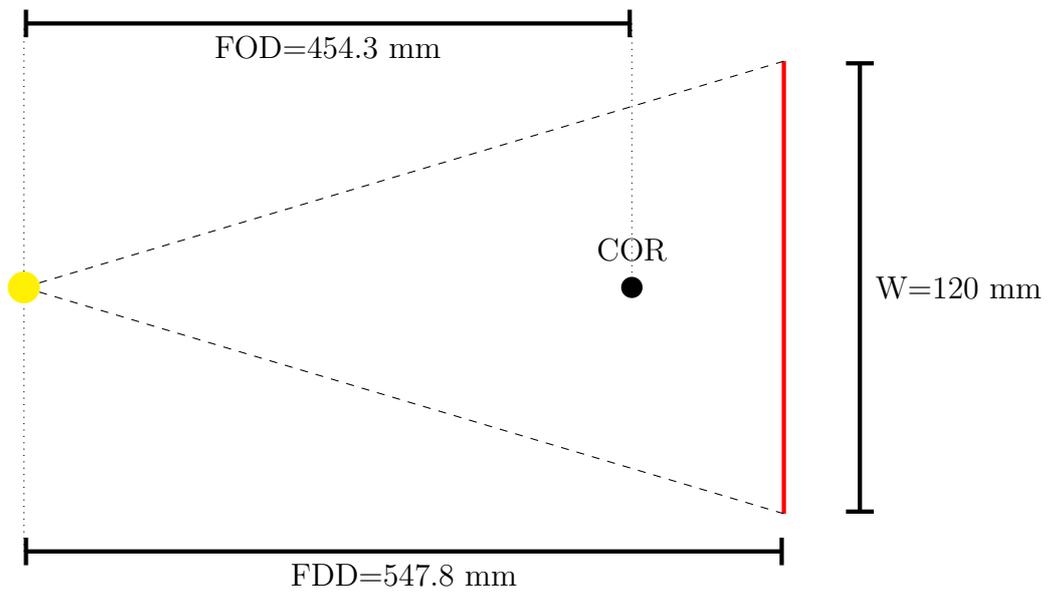
\begin{figure}
\begin{tikzpicture}[scale=2.0]
\draw [ultra thick,red] (2.5,-1.5) -- (2.5,1.5);
\draw[|-|, ultra thick] (-2.5,1.75) -- (1.5,1.75) node[below,midway]{FOD=454.3 mm}; 
\draw [|-|, ultra thick] (-2.5,-1.75) -- (2.5,-1.75)  node[below,midway]{FDD=547.8 mm}; 
\draw[|-|, ultra thick] (3.0,-1.5) -- (3.0,1.5); 
\draw[dotted] (-2.5,-1.75) -- (-2.5,1.75); 
\draw[dotted] (1.5,1.75) -- (1.5,0.0); 
\draw[dashed] (-2.5,0.0) -- (2.5,-1.5);
\draw[dashed] (-2.5,0.0) -- (2.5,1.5); 
\draw (3.65,0.0) node{W=120 mm}; 
\fill[thick] (1.5,0.0) circle (2pt);
\draw (1.5,0.25) node{COR};
\fill[thick, yellow] (-2.5,0.0) circle (3pt);
\end{tikzpicture}
\bigskip
\caption{Geometry of the measurement setup. Here FOD and FDD denote the focus-to-object distance and the focus-to-detector distance, respectively; the black dot COR is the center-of-rotation. The width of the detector (\ie{}, the red thick line) is denoted by W. The yellow dot is the X-ray source. To increase clarity, the $x$-axis and $y$-axis in this image are not in scale.}\label{k1}
\end{figure}

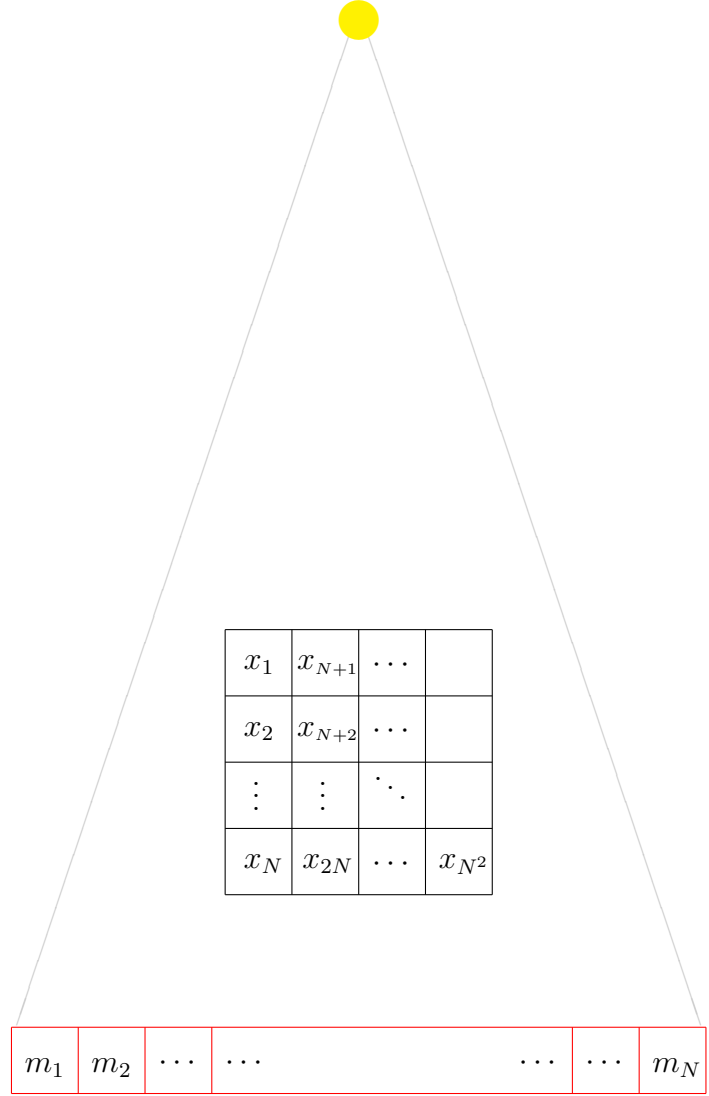
\begin{figure}
\begin{picture}(390,390)
\put(195,415){\color{lightgray}\line(1,-3){128}}
\put(195,415){\color{lightgray}\line(-1,-3){128}}

\put(195,410){\color{yellow}\circle*{25}}
\put(145,80){\line(1,0){100}}
\put(145,105){\line(1,0){100}}
\put(145,130){\line(1,0){100}}
\put(145,155){\line(1,0){100}}
\put(145,180){\line(1,0){100}}
\put(145,80){\line(0,1){100}}
\put(170,80){\line(0,1){100}}
\put(195,80){\line(0,1){100}}
\put(220,80){\line(0,1){100}}
\put(245,80){\line(0,1){100}}
\put(152,165){$x_1$}
\put(152,140){$x_2$}
\put(155,112){$\vdots$}
\put(152,90){$x_N$}
\put(172,165){$x_{\scriptscriptstyle N+1}$}
\put(172,140){$x_{\scriptscriptstyle N+2}$}
\put(180,112){$\vdots$}
\put(174,90){$x_{2N}$}
\put(200,164){$\cdots$}
\put(200,139){$\cdots$}
\put(200,115){$\ddots$}
\put(200,89){$\cdots$}
\put(225,90){$x_{N^2}$}

\put(65,5){\color{red}\line(1,0){260}}
\put(65,30){\color{red}\line(1,0){260}}
\put(65,5){\color{red}\line(0,1){25}}
\put(90,5){\color{red}\line(0,1){25}}
\put(115,5){\color{red}\line(0,1){25}}
\put(140,5){\color{red}\line(0,1){25}}
\put(275,5){\color{red}\line(0,1){25}}
\put(300,5){\color{red}\line(0,1){25}}
\put(325,5){\color{red}\line(0,1){25}}
\put(70,13){$m_1$}
\put(95,13){$m_2$}
\put(120,14){$\cdots$}
\put(145,14){$\cdots$}
\put(255,14){$\cdots$}
\put(280,14){$\cdots$}
\put(305,13){$m_N$}
\end{picture}
\caption{The organization of the pixels in the sinograms {\tt m}\,=\,$[m_1,m_2,\ldots,m_{60N}]^T$ and reconstructions {\tt x}\,=\,$[x_1,x_2,\ldots,x_{N^2}]^T$with $N=256$. The picture shows the organization for the first projection; after that in the full angular view case, the target takes $6$ degree steps counter-clockwise (or equivalently the source and detector take $6$ degree steps clockwise) and the following columns of {\tt m} are determined in an analogous manner.}\label{fig:pixelDemo}
\end{figure}

\begin{figure}
\includegraphics[scale=0.3]{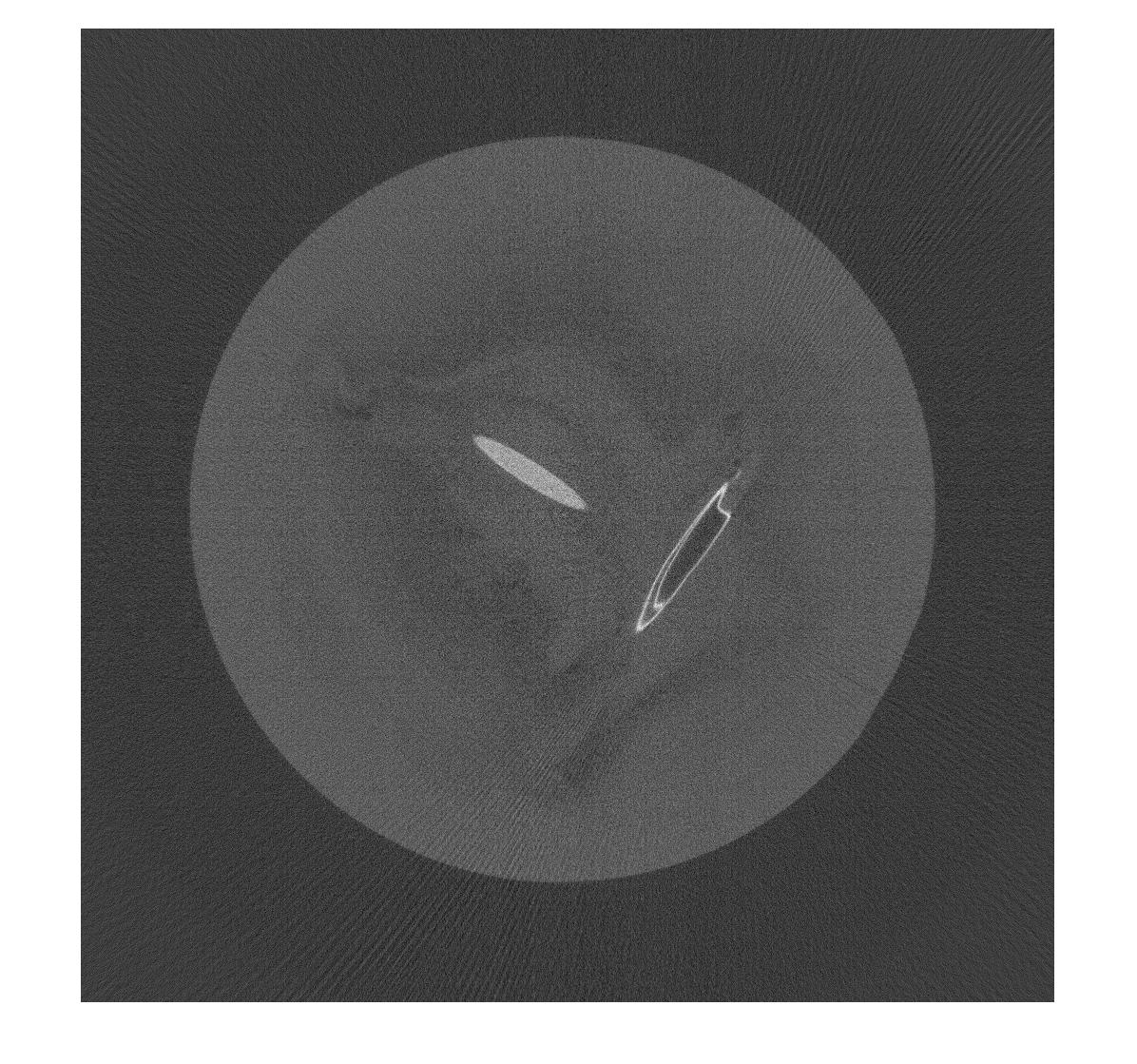}
\caption{ The high-resolution filtered back-projection reconstruction of the
time-dependent cross phantom computed from 360 projections.}\label{fig:GroundTruthReco}
\end{figure}
\bigskip
\clearpage\section{Example of using the data}\label{sec:Demo}

The following MATLAB code demonstrates how to use the data. The code is also provided as the separate MATLAB script file {\tt example.m} and it assumes the data files (or in this case at least the file {\tt DataStatic\_128x60.mat}) are included in the same directory with the script file.

\begin{verbatim}
% Load the measurement matrix and the sinogram from
% file DataStatic_128x60.mat
load DataStatic_128x60 A sinogram
m = sinogram;

% Compute a Tikhonov regularized reconstruction using
% conjugate gradient algorithm pcg.m
T		= 16; % number of time frames (16 or 30 depending on resolution)
N     = sqrt(size(A,2)/T);
alpha = 10; % regularization parameter
fun   = @(x) A.'*(A*x)+alpha*x;
b     = A.'*m(:);
x     = pcg(fun,b);

% Compute a Tikhonov regularized reconstruction from only
% 10 projections
[mm,nn] = size(m);
ind     = [];
for iii=1:nn/6
    ind = [ind,(1:mm)+(6*iii-6)*mm];
end
m2    = m(:,1:6:end);
A     = A(ind,:);
alpha = 10; % regularization parameter
fun   = @(x) A.'*(A*x)+alpha*x;
b     = A.'*m2(:);
x2    = pcg(fun,b);

% Take a look at the sinograms of the first slice and the reconstructions
figure
subplot(2,2,1)
imagesc(m(:,1:size(m,2)/T))
colormap gray
axis square
axis off
title('Sinogram, 60 projections')
subplot(2,2,3)
imagesc(m2(:,1:size(m2,2)/T))
colormap gray
axis square
axis off
title('Sinogram, 10 projections')
% reshape the reconstruction
x = reshape(x,N,N,T);
x2 = reshape(x2,N,N,T);

% Show the first slice reconstruction
subplot(2,2,2)
imagesc(imrotate(x(:,:,1),-98,'bilinear','crop'))
colormap gray
axis square
axis off
title({'Tikhonov reconstruction,'; '60 projections'})
subplot(2,2,4)
imagesc(imrotate(x2(:,:,1),-90,'bilinear','crop'))
colormap gray
axis square
axis off
title({'Tikhonov reconstruction,'; '10 projections'})
\end{verbatim}

\clearpage
\section{3D reconstruction}
The video below is three-dimensional "ground truth" reconstruction of the time-dependent cross phantom target. This reconstruction is made from the original measured data (360 angles) and reconstructed using the FDK-cuda algorithm of the ASTRA toolbox \footnote{The ASTRA Toolbox is open source under the GPLv3 license.} \cite{ASTRA_1} \cite{ASTRA_2}.

\bigskip
\bigskip
\includemedia[width=1\linewidth,height=1\linewidth,activate=pageopen,
passcontext,
transparent,
addresource=DynamicPhantom.mp4,
flashvars={source=DynamicPhantom.mp4}
]{\includegraphics[width=20\linewidth]{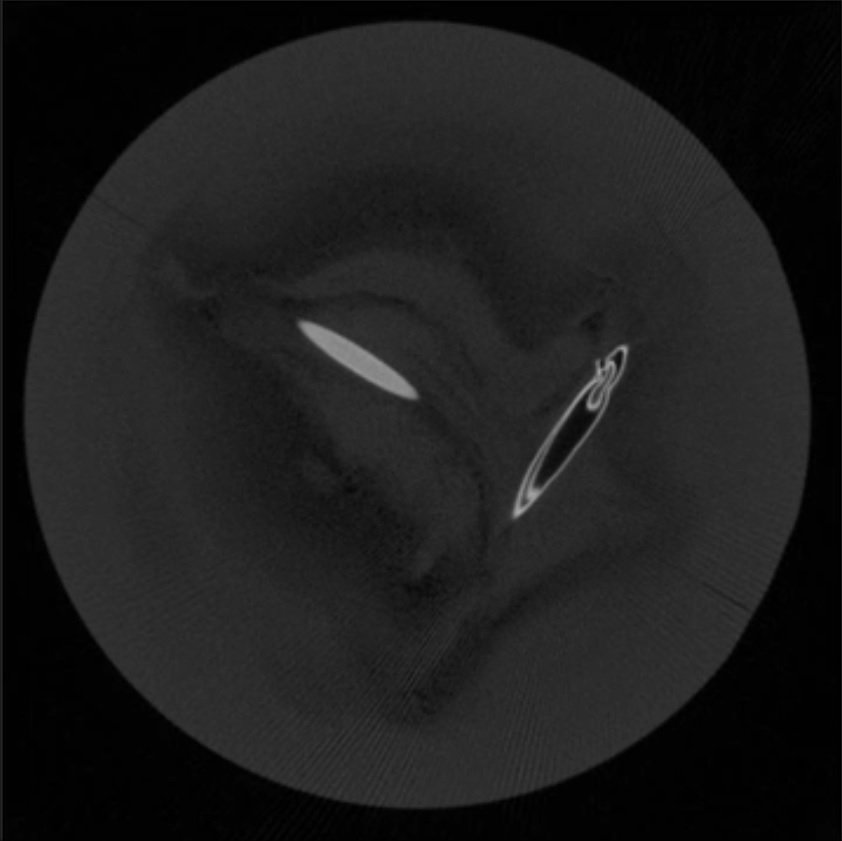}}{VPlayer.swf}

\end{document}